\documentclass{JHEP3} 
\usepackage{latexsym}

\newcommand{\cmu}{\check{\mu}}
\newcommand{\cnu}{\check{\nu}}
\newcommand{\crho}{\check{\rho}}
\newcommand{\nn}{\nonumber}
\newcommand{\bref}[1]{(\ref{#1})}
\def\tr{{\rm ~tr}\,}

\title{Towards the Super Yang-Mills Theory on the Lattice}

\author{Katsumi\ Itoh, Mitsuhiro\ Kato,$^a$~~Hideyuki\ Sawanaka,$^b$
Hiroto\ So$^b$ and Naoya\ Ukita$^b$\\
Faculty of Education, Niigata University, Niigata 950-2181, Japan\\
$^a$ Institute of Physics, University of Tokyo, Komaba, Meguro-ku, 153-8902 Tokyo, Japan\\
$^b$ Department of Physics, Niigata University, Niigata 950-2181,
Japan\\

	E-mail: \email{itoh@ed.niigata-u.ac.jp},
	\email{kato@hep1.c.u-tokyo.ac.jp},
	\email{hide@muse.hep.sc.niigata-u.ac.jp}, 
	\email{so@muse.hep.sc.niigata-u.ac.jp},
	\email{ukita@muse.hep.sc.niigata-u.ac.jp}}
\preprint{\heplat{0112052}\\	
NIIG-DP-01-7\\	  	
UT-Komaba/01-08}

\abstract{ We present an entirely new approach towards a realization of
 the supersymmetric Yang-Mills theory on the lattice.  The action
 consists of the staggered fermion and the plaquette variables
 distributed in the Euclidean space with a particular pattern.  The
 system is shown to have fermionic symmetries relating the fermion and
 the link variables.  }


\begin{document}
\section{Introduction}

 The vacuum structure of 4 dimensional super Yang-Mills (SYM) theory
has been studied by many authors after the works by Seiberg and
Witten\cite{SW}.\footnote{See \cite{Seiberg} for earlier works.}  In
their approach, dualities, the holomorphy and symmetries of low energy
theories play crucial roles.

The lattice gauge theory has proved to be a powerful formulation for
gauge theories and made definite success for non-supersymmetric
theories.  Many efforts have been paid to formulate the SYM theory on
the lattice with the expectation to provide a non-perturbative method.
However this hope has not been materialized yet due to various
difficulties against a realization of the SUSY algebra, such as: 1) a
realization of a massless Majorana fermion on the lattice; 2) the
absence of the Leibniz rule; 3) we may also count the absence of the
Bianchi identity in lattice formulations\cite{Bianchi}, which is crucial
to show the SUSY invariance of the continuum SYM.

Among various attempts to formulate the SYM on the lattice, here let us
recall a couple of them.  Curci and Veneziano\cite{cv} proposed a
formulation based on the Wilson fermion.  They claimed that the SUSY WT
identity as well as chiral WT identity would be fulfilled by the
fine-tuning of a fermion mass, ie, once the mass counter term is
appropriately chosen.  This claim has been confirmed at the one-loop
level in perturbation theory\cite{Taniguchi}, but there is no proof up
to now that it should hold in all orders.  In recent years, some
numerical studies have been performed for their approach\cite{Montovay}.
However the results are not so clearly favorable for the presence of
SUSY.  In this approach the fine-tuning is expected to produce a
massless Majorana fermion.

A latest attempt is based on the Domain Wall fermion (DWF)[7,8].  The
authors took the viewpoint that, except the gaugino mass term, any SUSY
breaking interaction is irrelevant around a SUSY fixed point when we
have the gauge symmetry and the right degrees of freedom.  To realize
the latter condition, the DWF was employed to produce a massless
Majorana fermion.  In this approach, a SUSY invariant fixed point is
assumed for the discussion of the continuum limit.  Its validity or
plausibility is yet to be clarified.  No evident relation is known
between the fixed point and the DWF approach.

Both of the above approaches use a ``mass'' term to remove the fermion
doublers.  If the SUSY requires that the gauge sector is influenced by
the fermion mass term, it may be hard to keep the gauge invariance.  So
we choose another way to avoid the doubling problem, ie, the staggered
fermion\cite{Susskind}.  It is not a new idea to use the staggered
fermion to realize the SUSY.  For earlier attempts, see
Refs. \cite{aratyn}.

In this paper, we propose an entirely new lattice theory with exact
fermionic symmetries which relate the gaugino and the gauge degrees of
freedom with Grassmann odd parameters.  An overview of its construction
and its properties will be described in the next section.  The rest of
this paper is organized as follows.  In the sections 3 and 4, we present
a model defined on a single hypercube, to be addressed as one cell
model, and show that it has an exact fermionic symmetry (preSUSY).  In
sections 5 and 6, the model is extended to the entire Euclidean space,
so that the preSUSY survives.  The last section is devoted to the
summary and discussion.

\section{An overview of our models}

It would be appropriate to present an overview of our approach and
strategy before going into the detailed explanation of our models.

We start our construction from a lattice theory defined on a single
hypercube.  This model will be addressed as the one cell model.  For
this model, written with link variables and site variables (real
fermions in the adjoint representation), we find an exact symmetry
which mixes the link and site variables. This symmetry will be called 
the preSUSY.

Although our theory contains the staggered fermion, it is different from
the earlier works listed in Ref. \cite{aratyn}.  Our model is defined on
a hypercube or a cell and its preSUSY transformation is clearly
different from the SUSY transformation in the latter approach.

The gauge part of the action $S_g$ is written by the ordinary plaquette
variables.  The fermion action $S_f$ consists of terms which corresponds
to fermions sitting at two neighboring sites connected by link
variables.  The coefficients for the terms in $S_f$ and those in the
preSUSY transformations on the link variables and fermions, denoted by
$\delta_U$ and $\delta_{\psi}$ respectively, are to be determined to
realize the preSUSY invariant action.  The preSUSY transformation of the
action consists of fermion cubed terms $\delta_U S_f$ and linear terms.
The vanishing of the former terms relate coefficients in $\delta_U$ and
$S_f$.  The condition allows us to have an interesting fermion action,
that is the staggered (Majorana) fermion coupled to link variables.
{}From the vanishing of the fermion linear terms we may find a condition
relating the remaining coefficient in $\delta_{\psi}$ to the rest.  In
order to keep the preSUSY invariance even at the quantum level, the path
integral measure should be invariant as well.  The Haar measure for the
link variables and the Grassmannian measure for the Majorana fermions is
to be invariant as a whole.  It will be found that the condition for
$\delta_U S_f =0$ is sufficient to show the invariance of the measure.

We extend the preSUSY invariant one cell model to the entire (Euclidean)
space and find an interacting cell model.  A naive attempt to extend the
model tends to end up with the uninteresting ``SUSY'' transformations:
in the naive continuum limit they become $O(a)$, thus the symmetry is a
lattice artifact.  The extension presented here avoids the above
mentioned difficulty owing to a non-trivial space structure. 

The fermion part of the action $S_f$ is extended for all the links.  The
non-trivial structure is introduced for the gauge part.  We classify the
hypercubes in the entire space into three categories, the E-cells,
O-cells and the rest.  The two dimensional example is shown in Fig. 3.
We introduce the plaquette variables only for E- and O-cells.  Those for
other hypercubes are missing.  A link variable or a plaquette variable
is associated with either an E-cell or O-cell, while the fermion as a
site variable belongs to both E- and O-cells.  Accordingly each term in
the action belongs to either an E- or an O-cell, and the interactions
between cells are present due to the two-sided nature of the fermions.
The discrete translational invariance by $2a$ is present by
construction.  The condition of the preSUSY invariance of the action
relates coefficients defined for neighboring cells.  Due to this
condition, the number of independent parameters in the preSUSY
transformation are reduced drastically.

\section{One cell model}

Consider a cell or hypercube in D dimensional Euclidean space.  The
coordinates of the sites may be written as $r \equiv (r_1, r_2, \cdots)$
with $r_i = 0$ or $1$.  In this paper, we set the lattice spacing as
$a=1$.

The gauge action on this cell is given as
\begin{eqnarray}
S_{g}= -\beta\sum_{0<\mu<\nu} \sum_{n=r(\cmu\cnu)}
\tr \left(U_{n,\mu\nu} +  U_{n,\nu\mu}\right)
\label{S g}
\end{eqnarray}
where $r(\cmu\cnu)$ implies sites with $r^{\mu}=r^{\nu}=0$.
$U_{n,\mu\nu}$ is a plaquette variable which goes first to the $\mu$
direction and then to the $\nu$ direction starting from the site $n$.
In order for the plaquette to be on the cell, the starting site $n$ is
restricted to that with $r^{\mu}=r^{\nu}=0$.

Our action of the real staggered fermion is
\begin{eqnarray}
 S_{f}=\sum_{0<\rho} \sum_{n=r(\crho)}b_{\rho}(n) 
\tr \left(
\psi_{n}U_{n,\rho}\psi_{n+\hat{\rho}}U_{n,\rho}^{\dagger}\right)  
\equiv \sum_{0<\rho} \sum_{n=r(\crho)}b_{\rho}(n) 
\tr \left(
\psi_{n}\xi_n^{\rho}\right).  
\label{S f}
\end{eqnarray}
Here the coefficient $b_{\mu}(n)$ is a sign factor for the staggered
fermion.  We also introduced new notations $r(\crho) \equiv (r_1, \cdots ,
r_{\rho-1},r_{\rho}=0,r_{\rho+1},\cdots)$ and $\xi_n^{\rho} \equiv
U_{n,\rho} \psi_{n+{\hat \rho}}U_{n,\rho}^{\dagger}$.
It is convenient
to treat two fermions in \bref{S f} in a symmetric manner.  So let us
write the fermion action as
\begin{eqnarray}
S_f =\sum_{0<\rho} \sum_{n=r(\crho)}b_{-\rho}({n+{\hat \rho}}) \tr \left(\psi_{n+\hat{\rho}}U_{n,\rho}^{\dagger}\psi_{n}U_{n,\rho}\right) \equiv \sum_{0<\rho} \sum_{n=r(\crho)}b_{-\rho}(n+{\hat \rho}) \tr \left(\psi_{n+{\hat \rho}}\xi_{n+{\hat \rho}}^{-\rho}\right).  
\nonumber
\end{eqnarray}
Here $b_{-\rho}(r(\crho)+{\hat \rho})\equiv - b_{\rho}(r(\crho))$ and
$\xi_{n+{\hat \rho}}^{-\rho} \equiv U_{n,\rho}^{\dagger}\psi_n
U_{n,\rho}$.  
Later we extend the fermion action to the entire Euclidean space.  The 
sign factor $b_{\mu}(n)$ should be properly defined in the whole space 
so that the action describes the staggered fermion.

We assume the form of preSUSY transformation and find conditions among
coefficients required for the invariance of the action.  Let us write
down our ansatz for the preSUSY transformation.  For the link variable, it
is
\begin{eqnarray}
\delta U_{n,\mu} =
(\alpha \cdot \xi)_{n,\mu}U_{n,\mu}
+U_{n,\mu}(\alpha \cdot \xi)_{n+{\hat \mu},\mu}
\label{One cell del U}
\end{eqnarray}
where 
\begin{eqnarray}
(\alpha \cdot \xi)_{n,\mu}\equiv \sum_{\rho} 
\alpha_{n,\mu}^{\rho[n]}\xi_n^{\rho[n]}
\label{alpha dot xi}
\end{eqnarray}
and $\rho[n] \equiv (-1)^{n_{\rho}} {\rho}$.  The transformation
parameters $\alpha_{n,\mu}^{\rho[n]}$ are Grassmann odd.  The preSUSY
transformation of fermion fields is assumed to be,
\begin{eqnarray}
\delta \psi_{n} =  \frac{1}{2} \sum_{0 < \mu,\nu}
C_{n}^{(\mu\nu)[n]} \left(
U_{n,(\mu\nu)[n]} - 
U_{n,(\nu\mu)[n]}\right), 
\label{del f}
\end{eqnarray}
where ${(\mu\nu)[n]} \equiv {\mu[n]}{\nu[n]}$.  The parameters
$C_{n}^{(\mu\nu)[n]}$ are also Grassmann odd and antisymmetric under the
exchange of $\mu$ and $\nu$.

Note that coefficients $C_{n}^{(\mu\nu)[n]}$ and $\alpha_{n,\mu}^{\nu[n]}$
have indices referring to a site $n$ and the face defined by two
directions, $\mu$ and $\nu$.  So for a given site and a given face, we
introduce four independent parameters by $\alpha_{n,\mu}^{\nu[n]}$ while we
do only one by $C_{n}^{(\mu\nu)[n]}$.  These counting will be useful for
later discussion.  From now on we may call these parameters 
$\alpha$-parameters and $C$-parameters for convenience.

Some explanations are in order on our ansatz for the preSUSY
transformation.  The form of fermion transformation in \bref{del f} is
motivated by the continuum SUSY transformation, where the
fermion is related to the field strength with its Lorentz indices
contracted with the gamma matrices.  In order to cancel terms produced by
the fermion transformation acting on $S=S_g+S_f$, we chose eq. \bref{One
cell del U} as the transformation of the link variable.  Naturally, we
expect that the coefficients of transformations tend to the gamma
matrices in a continuum limit: $C_{n}^{(\mu\nu)[n]} \propto
\gamma^{[\mu}\gamma^{\nu]}$ and $\alpha_{n,\mu}^{\nu} \propto
\gamma^{\mu} $.

In the next section, we derive relations between the coefficients in the
action and the parameters in the preSUSY transformation by requiring the
invariance of the action.  We will find that the relations may be solved
for any given coefficients in the action, $\beta$ and $b_{\mu}(n)$.  So
let us assume that these are appropriately chosen (non-vanishing)
coefficients.

\section{preSUSY invariance of one cell model}

We derive relations on the coefficients in our action and preSUSY
transformation.  The transformation of the action consists of three
terms,
\begin{eqnarray}
\delta S = \delta_U S_g + \delta_U S_f + \delta_{\psi} S_f.
\label{delS}
\end{eqnarray} 
Here $\delta_U$ and $\delta_{\psi}$ denote the transformations of link
variables and fermions given in eqs. \bref{One cell del U} and \bref{del
f}, respectively.  The first term of \bref{delS} consists of fermion
cubed terms and the rest linear terms.  So they must vanish separately,
$\delta_U S_f=0$ and $\delta_U S_g + \delta_{\psi} S_f =0$.  We also
study the invariance of the path integral measure.

After a straightforward but tedious calculations, we find that the
condition $\delta_U S_f =0$ gives us
\begin{eqnarray}
(-)^{r_{\nu}}\frac{\alpha_{r,\mu}^{\nu[r]}}{b_{\nu}(r)}
+
(-)^{r_{\mu}}\frac{\alpha_{r,\nu}^{\mu[r]}}{b_{\mu}(r)}=0.
\label{One cell cubed}
\end{eqnarray}
In Fig.1, we present a graphical representation of a term in $\delta_U S_f$.
Two terms of eq. \bref{One cell cubed} correspond to two different ways
to obtain this graph.

\begin{picture}(100,100)
\thicklines
\put(60,40){\circle{6}}
\put(60,85){\circle{6}}
\put(105,40){\circle{6}}
\put(40,60){\Large $\nu$}
\put(80,25){\Large $\mu$}
\put(48,28){\Large $r$}
\put(60,43){\line(0,1){39}}
\put(63,40){\line(1,0){39}}
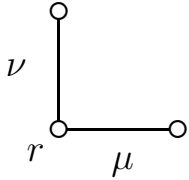
\put(10,10){\small Figure 1: A graphical representation of a term in
 $\delta_U S_f$.  The fermions}
\put(56,-2){\small are located at sites indicated by circles.}
\end{picture}
\vspace{3mm}

Let us consider its implication.  Eq.~\bref{One cell cubed} simply tells
us that the combination
$(-)^{r_{\nu}}{\alpha_{r,\mu}^{\nu[r]}}/{b_{\nu}(r)}$ is
anti-symmetric under the exchange of $\mu$ and $\nu$.  In
particular, the relation with $\mu=\nu$ leads to the vanishing of the
diagonal element of $\alpha$-parameter.  Thus we learn from
eq. \bref{One cell cubed} that there is only one independent element in
$\alpha_{r,\mu}^{\nu[r]}$ for the site $r$ and the face determined by
$\mu$ and $\nu$.

The other condition $\delta_U S_g + \delta_{\psi} S_f =0$ contains new
transformation parameters $C_r^{(\mu\nu)[r]}$.  We find two different
relations,
\begin{eqnarray}
b_{\rho}(r) C_r^{(\mu\nu)[r]} &=& \beta \Bigl((-)^{r_{\mu}}\alpha_{r,\mu}^{\rho[r]}- 
(-)^{r_{\nu}}\alpha_{r,\nu}^{\rho[r]}\Bigr),
\label{One cell linear 1}\\
b_{\mu}(r)C_r^{(\mu\nu)[r]}&+&b_{\nu}(r_d)C_{r_d}^{(\mu\nu)[r_d]}=-\beta 
\Bigl(
(-)^{r_{\nu}}\alpha_{r,\nu}^{\mu[r]}+(-)^{r_{\mu}}\alpha_{r_d,\mu}^{\nu[r_d]}
\Bigr).
\label{One cell linear 2}
\end{eqnarray}
In eqs. \bref{One cell linear 1} and \bref{One cell linear 2}, the lhs (rhs)
comes from $\delta_{\psi}S_f$ ($\delta_U S_g$).  

The fermion action \bref{S f} consists of terms with two $\psi$
connected by link variables.  Since the transformation \bref{del f}
replaces one of $\psi$ by a plaquette variable, generically the
corresponding term may be expressed by a three dimensional figure shown
in Fig. 2(a).
In $\delta_U S_g + \delta_{\psi} S_f$, the terms with $\rho \ne \mu,
\nu$ give us the relation \bref{One cell linear 1}.  When $\rho=\mu$ or
$\nu$, we have to look at the cancellation condition with some care.  As
shown in Fig. 2(b), the figure becomes two dimensional and the same
figure may be produced by the transformation of fermions located at $r$
as well as $r_d$.  Here $r_d \equiv (r_1, r_2, \cdots, 1-r_{\mu},
\cdots, 1-r_{\nu}, \cdots)$ is the site diagonal to $r$ on the $\mu\nu$
face including the site $r$.  We find eq. \bref{One cell linear 2}, in
which coefficients associated with two sites $r$ and $r_d$ are related.

\vspace{3mm}
\begin{picture}(360,140)
 \thicklines
\put(30,70){\circle*{8}}
\put(30,130){\circle{8}}
\put(30,30){(a) $\rho \ne \mu, \nu$}
\put(10,130){\Large $\psi$}
\put(15,55){\Large $r$}
\put(15,100){\Large $\rho$}
\put(60,55){\Large $\mu$}
\put(120,80){\Large $\nu$}
\put(30,70){\line(0,1){56}}
\put(30,70){\line(1,0){70}}
\put(30,70){\line(2,3){30}}
\put(100,70){\line(2,3){30}}
\put(60,115){\line(1,0){70}}
\put(200,30){(b) $\rho = \mu$ or $\nu$}
\put(300,115){\circle*{8}}
\put(200,70){\circle*{8}}
\put(270,70){\circle{8}}
\put(185,55){\Large $r$}
\put(310,125){\Large $r_d$}
\put(276,54){\Large $\psi$}
\put(230,55){\Large $\mu$}
\put(290,80){\Large $\nu$}
\put(200,70){\line(1,0){66}}
\put(200,70){\line(2,3){30}}
\put(273,73){\line(2,3){27}}
\put(230,115){\line(1,0){70}}
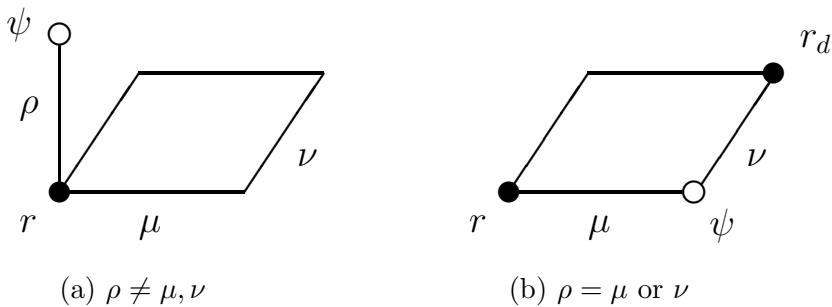
\put(20,10){\small Figure 2: Graphical representations of terms in
$\delta_U S_f +\delta_{\psi} S_f$.}
\end{picture}

Using \bref{One cell cubed}, we may rewrite \bref{One cell linear 1} as
\begin{eqnarray}
C_r^{(\mu\nu)[r]}=\beta (-)^{1+r_{\rho}}
\Bigl[
\frac{\alpha_{r,\rho}^{\mu[r]}}{b_{\mu}(r)}
-\frac{\alpha_{r,\rho}^{\nu[r]}}{b_{\nu}(r)}
\Bigr]. 
\label{C by alpha}
\end{eqnarray}
In eq. \bref{C by alpha}, the index $\rho$ on the rhs could take any
value. Choosing $\rho = \mu$ and $\nu$, we find the equalities,
\begin{eqnarray}
C_r^{(\mu\nu)[r]}=\beta (-)^{r_{\mu}} \frac{\alpha^{\nu[r]}_{r,\mu}}{b_{\nu}(r)}
=-\beta (-)^{r_{\nu}} \frac{\alpha^{\mu[r]}_{r,\nu}}{b_{\mu}(r)}.
\label{C vs alpha}
\end{eqnarray}
Eq. \bref{One cell linear 2} is found to produce the same relations and
does not carry any further information.

{}From eq. \bref{C vs alpha}, we learn that the $C$-parameters are
written by the $\alpha$-parameters or vice versa.  By combining this
with the result in the previous section, we find that there is one
independent parameter for a site and a face.  So there are ${}_D{\rm
C}_2$ independent parameters for a site.  However this counting is not
sufficient since a further restriction is present.  From \bref{C by
alpha} we easily find the cyclic relation
\begin{eqnarray}
C_r^{(\mu\nu)[r]}+C_r^{(\nu\lambda)[r]}+C_r^{(\lambda\mu)[r]}=0
\label{cyclic relation}
\end{eqnarray}
for any combination of $\mu$, $\nu$ and $\lambda$.   

Eq. \bref{cyclic relation} is a ``local'' relation; it holds at each
site.  So let us consider it at the origin for convenience.
By ignoring the site index, it is written simply as
\begin{eqnarray}
C^{\mu\nu}+C^{\nu\lambda}+C^{\lambda\mu}=0.
\label{cyclic relation2}
\end{eqnarray}
Observing that $C$ are antisymmetric with respect to the upper indices,
we use the following analogy to find the number of independent $C$.
Take D independent points in D dimensional (or larger) space and name
them as $1,~2,~3,~\cdots$.  Consider the vectors connecting the points,
$\overrightarrow{12}=-\overrightarrow{21}$ etc.  There are ${}_D{\rm
C}_2$ of them altogether.  Obviously, these vectors satisfy the relation
similar to \bref{cyclic relation2}: the cyclic relation simply implies
that three vectors connecting three points form a triangle.  Including
the antisymmetric property, we may identify $C^{\mu\nu}$ and
$\overrightarrow{\mu\nu}$ for solving eq. \bref{cyclic relation2}.  From
this identification, we clearly see that there are $D-1$ independent
$C$-parameters at a site.

Before closing this section we would like to confirm the invariance of
the path integral measure at the first order of the transformation
parameters.  We will soon find that there appears no new relation from
the invariance.  Actually the vanishing of the diagonal elements of the
$\alpha$-parameter, concluded from eq. \bref{One cell cubed}, is enough
to show the invariance of the path integral measure.

Let us express eqs. \bref{One cell del U} and \bref{del f} schematically 
as
\begin{eqnarray}
\delta U &=& F(\alpha,\psi,U),\nonumber\\
\delta \psi &=& G(C, U).\nonumber
\end{eqnarray}
{}For the present discussion, we will see it convenient to achieve the
transformation by taking the following two steps,
\begin{eqnarray}
\left(
  \begin{array}{c}
   U \\ \psi
  \end{array}
\right)
\rightarrow
\left(
  \begin{array}{c}
   U+F(\alpha,\psi-\delta \psi,U) \\ \psi
  \end{array}
\right)
\rightarrow
\left(
  \begin{array}{c}
   U+F(\alpha,\psi,U) \\ \psi+\delta \psi 
  \end{array}
\right).
\end{eqnarray}
The path integral measure may be written schematically as $\prod d\psi
\prod d U$.  The Haar measure $d U$ for each link variable is invariant
under the action of a unitary matrix from the left (or the right).  In the
second step, we transform only the fermions.  The path integral measure
is obviously invariant in this step.  So we are left to study the first
step.  In the lowest order in the transformation parameters, it is the
transformation only of the link variables, $\delta U =
F(\alpha,\psi,U)$.  It is simply the infinitesimal form of unitary
transformations acted both from the left and right, if $(\alpha \cdot
\xi)$ is pure imaginary and it does not contain the transformed variable
$U_{n,\mu}$ itself.  The latter condition may be expressed as
\begin{eqnarray*}
\alpha^{\mu}_{n,\mu}=\alpha^{-\mu}_{n+{\hat \mu},\mu}=0,
\end{eqnarray*}
the vanishing of the diagonal elements.

\section{Interacting cell model}

Although we showed that our one cell model is invariant under the
Grassmannian transformation, it is highly non-trivial whether we may
extend it to the entire D-dimensional space.  Here we present one
successful way of its extension.

Take a site with all its coordinate elements as even (or odd) integers.  This
site will be addressed as an even (or odd) reference point and its
coordinate is indicated by $N$.  Starting from a reference point, we may
form a cell, a hypercube, with unit vectors towards positive directions.
A cell formed from an even (or odd) reference point is called  an
E-cell (O-cell).  The two dimensional example is a traditional
pattern\footnote{This is called  ``Ichimatsu pattern'' in the Japanese
tradition.}  shown in Fig. 2.  We put copies of the one cell model on
the E-cells as well as on the O-cells.  Owing to our restriction on the
reference points, there appear the spaces without the cell structure.

The link variables belong either to an E-cell or to an O-cell, while a
fermion is on the site and associated with a pair of neighboring E-cell
and O-cell.  The action of this interacting cell model is a simple
extension of the one cell model.  The interaction between cells are
described solely by the fermion action.  The link variables in a pair of
neighboring cells interact through the fermion located at the shared
site.  To be described below, we modify our ansatz for the preSUSY
transformation by including the contributions from E-cells and O-cells.

The sites on a cell have the coordinates denoted as $n=N+r$.  Here $r$
is the relative coordinate from the reference point $N$: the component
of $r$ is either $1$ or $0$.  As for the one cell model, we denote by
$r(\cmu)$ a relative coordinate with $r_{\mu}=0$; similarly
$r(\cmu\cnu)$ denotes a relative coordinate with $r_{\mu}=0$ and
$r_{\nu}=0$.

A site $n$ belongs to two cells and its coordinate may be written in two
different ways,
\begin{eqnarray}
n= N + r = N^{\prime} + r^{\prime},
\end{eqnarray}
where $N^{\prime} \equiv N - e + 2r$ and $r^{\prime} \equiv e-r$ with $e
\equiv (1,1,1,1,\cdots)$.  It is easy to confirm that $N$ and
$N^{\prime}$ are the reference points of two neighboring cells.  In the
following we use the notations $n \equiv N + r$ and $n^{\prime} \equiv
N^{\prime} + r^{\prime}$ to represent the same site but in reference to
two different cells.

\vspace{-2mm}
\begin{picture}(300,180)
 \put(30,30){\framebox(30,30){}}
 \put(94,30){\framebox(30,30){}}
 \put(158,30){\framebox(30,30){}}
 \put(30,94){\framebox(30,30){}}
\put(93,93){\circle*{6}}
 \put(94,94){\framebox(30,30){O-cell}}
 \put(80,99){\large $P$}
 \put(158,94){\framebox(30,30){}}
\thicklines
 \put(62,62){\framebox(30,30){E-cell}}
 \put(126,62){\framebox(30,30){}}
 \put(190,62){\framebox(30,30){}}
 \put(62,126){\framebox(30,30){}}
 \put(126,126){\framebox(30,30){}}
 \put(190,126){\framebox(30,30){}}
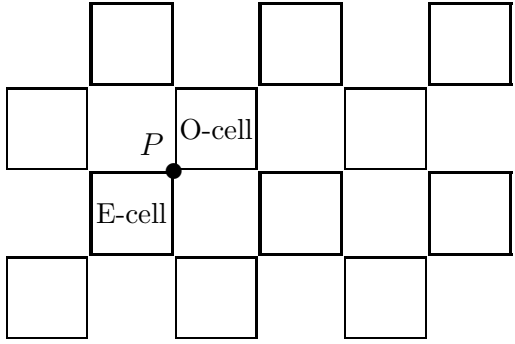
\put(20,12){\small Figure 3: Ichimatsu pattern. 
The thick (thin) cells are E-(O-)cells.}
\put(66,0){\small The point P is shared by two neibouring cells.}
\end{picture}
\vspace{3mm}

The action for the interacting model is easily obtained from \bref{S g}
and \bref{S f}.  The generic notation for the site $n$ is now to be
understood as the sum of the reference and the relative coordinates,
$n=N+r$.  The action is 
\begin{eqnarray}
S_{g}&=& -\beta \sum_N \sum_{0<\mu<\nu} \sum_{n=N+r(\cmu\cnu)}
\tr \left(U_{n,\mu\nu} +  U_{n,\nu\mu}\right),
\label{S g 2}\\
S_{f}&=&\sum_N \sum_{0<\rho} \sum_{n=N+r(\crho)}b_{\rho}(n) 
\tr \left(
\psi_{n}U_{n,\rho}\psi_{n+\hat{\rho}}U_{n,\rho}^{\dagger}\right).  
\label{S f 2}
\end{eqnarray}

The preSUSY transformation is modified as well.  On the link variable, it
takes the same form as the one cell model given in eq. \bref{One cell del U},
but now $(\alpha \cdot \xi)_{n,\mu}$ on the rhs is to be understood as
follows,
\begin{eqnarray}
(\alpha \cdot \xi)_{n,\mu}= 
\sum_{\rho} \Bigl(\alpha_{n,\mu}^{\rho[r]}\xi_n^{\rho[r]}
+ {\tilde \alpha}_{n,\mu}^{-\rho[r]}\xi_n^{-\rho[r]}\Bigr).
\label{alpha dot xi 2}
\end{eqnarray}
{}For the one cell model, $\rho[n]=(-)^{n_{\rho}}\rho$ is chosen so that
the fermion in $\xi_n^{\rho[n]}$ stays inside the cell.  So the fermions
in the second term of \bref{alpha dot xi 2} are in the neighboring cell.
A new set of ${\tilde \alpha}$-parameters are introduced accordingly.

Since a fermion variable is associated with two neighboring cells, a
natural extension of eq. \bref{del f} is to include plaquette variables
of those cells.  We modify the transformation accordingly,
\begin{eqnarray}
\delta \psi_{n} &=&  \sum_{0 < \mu < \nu}
\Bigl[C_{n}^{(\mu\nu)[r]} \left(
U_{n,(\mu\nu)[r]} - 
U_{n,(\nu\mu)[r]}\right)\Bigr]_{n=N+r}\nn\\
&+&\sum_{0 < \mu < \nu}
\Bigl[C_{n^{\prime}}^{(\mu\nu)[r^{\prime}]} \left(
U_{n^{\prime},(\mu\nu)[r^{\prime}]} - 
U_{n^{\prime},(\nu\mu)[r^{\prime}]}\right)\Bigr]_{n^{\prime}=N^{\prime}+r^{\prime}}.
\label{Interacting cell del psi}
\end{eqnarray}

\section{preSUSY invariance of interacting cell model}

We take the variation of the action given in \bref{S g 2} and \bref{S f
2} under the modified preSUSY transformation.  From now on, in writing
down the cancellation conditions, let us use $n=N+r$ and $n^{\prime}
\equiv N^{\prime} + r^{\prime}$ to represent the coordinates in the
E-cell and O-cell, respectively.

Consider the fermion cubed terms in $\delta S$, or $\delta_U S_f$.  Even
in the interacting cell models, the terms have the same graphical
representation as shown in Fig. 3.  Imagine to draw the diagram on the
pattern in Fig. 2.  We may put it within a single cell or over a pair of 
neighboring cells.  When it is drawn within a single cell,
the corresponding vanishing condition is the same as the one
cell model.  The condition is on the $\alpha$-parameters in the first
term of eq. \bref{alpha dot xi 2}.
\begin{eqnarray}
(-)^{r_{\nu}}\frac{\alpha_{n,\mu}^{\nu[r]}}{b_{\nu}(n)}
+
(-)^{r_{\mu}}\frac{\alpha_{n,\nu}^{\mu[r]}}{b_{\mu}(n)}=0,
\nonumber\\
(-)^{r^{\prime}_{\nu}}\frac{\alpha_{n^{\prime},\mu}^{\nu[r^{\prime}]}}{b_{\nu}(n^{\prime})}
+
(-)^{r^{\prime}_{\mu}}\frac{\alpha_{n^{\prime},\nu}^{\mu[r^{\prime}]}}{b_{\mu}(n^{\prime})}=0
.
\label{interacing cell cubed}
\end{eqnarray}
When we put the diagram in Fig. 1 over a pair of neighboring cells, we
find relations relating newly introduced ${\tilde \alpha}$-parameters.
\begin{eqnarray}
(-)^{r_{\mu}}b_{\mu}(n){\tilde \alpha}_{n,\mu}^{-\nu[r]}
+
(-)^{r^{\prime}_{\nu}}b_{\nu}(n^{\prime})
{\tilde \alpha}_{n^{\prime},\nu}^{-\mu[r^{\prime}]}=0.
\label{fermion cubed 2}
\end{eqnarray}
Eq. \bref{fermion cubed 2} relates the ${\tilde \alpha}$-parameters from
two neighboring cells and reduces the independent parameters to the half.

As for the fermion linear condition, we again have the same conditions
as the one cell model,
\begin{eqnarray}
C_n^{(\mu\nu)[r]}=\beta (-)^{r_{\mu}} 
\frac{\alpha^{\nu[r]}_{n,\mu}}{b_{\nu}(n)}
=-\beta (-)^{n_{\nu}} 
\frac{\alpha^{\mu[r]}_{n,\nu}}{b_{\mu}(n)},\nonumber\\
C_{n^{\prime}}^{(\mu\nu)[{r^{\prime}}]}=\beta (-)^{{r^{\prime}}_{\mu}} 
\frac{\alpha^{\nu[{r^{\prime}}]}_{{n^{\prime}},\mu}}{b_{\nu}({n^{\prime}})}
=-\beta (-)^{{r^{\prime}}_{\nu}} 
\frac{\alpha^{\mu[{r^{\prime}}]}_{{n^{\prime}},\nu}}{b_{\mu}({n^{\prime}})}.
\label{C vs alpha 2}
\end{eqnarray}
The remaining conditions from $\delta_U S_f + \delta_{\psi} S_f =0$
relate the $C$-parameters to the new ${\tilde \alpha}$-parameters
\begin{eqnarray}
b_{\rho}(n) C_{n^{\prime}}^{(\mu\nu)[r^{\prime}]}
&=& \beta \Bigl(
(-)^{r_{\mu}^{\prime}} {\tilde \alpha}_{n^{\prime},\mu}^{-\rho[r^{\prime}]}-
(-)^{r_{\nu}^{\prime}} {\tilde \alpha}_{n^{\prime},\nu}^{-\rho[r^{\prime}]}
\Bigr),
\label{Interacting cell linear 1}\\
b_{\rho}(n^{\prime}) C_{n }^{(\mu\nu)[r]}
&=& \beta \Bigl(
(-)^{r_{\mu} } {\tilde \alpha}_{n ,\mu}^{-\rho[r]}-
(-)^{r_{\nu} } {\tilde \alpha}_{n ,\nu}^{-\rho[r]}
\Bigr).
\label{Interacting cell linear 2}
\end{eqnarray}
In the above we wrote two almost the same equations.  They differ only in 
the indices.  Recall that the indices with a prime is associated with
the O-cell, those without a prime is for the neighboring E-cell.  

Now let us count the number of independent parameters for the
interacting cell model.  We may rewrite eq. \bref{Interacting cell
linear 1} for ${\tilde \alpha}_{n^{\prime},\mu}^{-\rho[r^{\prime}]}$.
Thus the parameter with different lower index $\mu$ are related each
other.  From \bref{Interacting cell linear 2} we see the same holds for
${\tilde \alpha}_{n, \mu}^{-\rho[r]}$.  Taking into account of the fact
that the $\alpha$-parameters for the neighboring cells are related by
\bref{fermion cubed 2}, the above stated observations imply that there
is only one independent ${\tilde \alpha}$-parameter at the site, say
${\tilde \alpha}_{n^{\prime},1}^{-1[r^{\prime}]}$.

Including the consideration for the one cell model, we conclude that
there are $2D-1$ independent parameters at each site: $D-1$
$C$-parameters from two neighboring cells and plus one ${\tilde
\alpha}$-parameter just stated.  Let us write the rest of parameters in
terms of the independent parameters, $C_n^{(\mu 1)[r]},
C_{n^{\prime}}^{(\mu 1)[r^{\prime}]}$ and ${\tilde
\alpha}_{n^{\prime},1}^{-1[r^{\prime}]}$,
\begin{eqnarray}
C_n^{(\mu \nu)[r]} &=& C_n^{(\mu 1)[r]}-C_n^{(\nu 1)[r]},
\nonumber\\
C_{n^{\prime}}^{(\mu \nu)[r^{\prime}]}
&=& C_{n^{\prime}}^{(\mu 1)[r^{\prime}]} 
- C_{n^{\prime}}^{(\nu 1)[r^{\prime}]},
\nonumber\\
\alpha_{n,\mu}^{\nu[r]}&=& (-)^{r_{\mu}} \frac{b_{\nu}(n)}{\beta}
\left( C_n^{(\mu 1)[r]}-C_n^{(\nu 1)[r]}\right),
\nonumber\\
\alpha_{{n^{\prime}},\mu}^{\nu[r^{\prime}]}&=& (-)^{r^{\prime}_{\mu}} \frac{b_{\nu}(n)}{\beta}
\left( C_{n^{\prime}}^{(\mu 1)[r^{\prime}]}-C_{n^{\prime}}^{(\nu 1)[r^{\prime}]}\right)
\label{explicit solution},\\
{\tilde \alpha}_{n,\mu}^{-\nu[r]}&=& 
- \frac{b_{\nu}(n^{\prime})}{b_1(n)}(-)^{r_1^{\prime}+r_{\mu}}
{\tilde \alpha}_{n^{\prime},1}^{-1[r^{\prime}]}
+(-)^{r_{\mu}} \frac{b_{\nu}(n^{\prime})}{\beta}
\left( C_n^{(\mu 1)[r]}-C_{n^{\prime}}^{(\nu 1)[r^{\prime}]}\right),
\nonumber\\
{\tilde \alpha}_{n^{\prime},\mu}^{-\nu[r^{\prime}]}&=& 
\frac{b_{\nu}(n)}{b_1(n)}(-)^{r_1^{\prime}+r^{\prime}_{\mu}}
{\tilde \alpha}_{n^{\prime},1}^{-1[r^{\prime}]}
+(-)^{r^{\prime}_{\mu}} \frac{b_{\nu}(n)}{\beta}
\left( C_{n^{\prime}}^{(\mu 1)[r^{\prime}]}-C_n^{(\nu 1)[r]}\right).
\nonumber
\end{eqnarray}

A comment is in order on the invariance of the path integral measure.
It is easy to realize that the problem reduces to that for the one cell
model.  Note that the new terms in the transformation in
eqs. \bref{alpha dot xi 2} relate the variables in different cells.  So
there is no new contribution to the jacobian beyond the one cell model.

\section{Summary and Discussion}

In this paper we presented a lattice model with an exact fermionic
symmetry.  By requiring the invariance of the action, we derived
relations among the coefficients in the action and the parameters in the
preSUSY transformation.  We found a finite number of independent
transformation parameters on each site by solving the relations.  The
path integral measure is found to be invariant if these relations are
satisfied.  Note that, in solving the relations, no assumption was made
for the coefficients in the action.  In particular we did not need an
expression of $b_{\mu}(n)$ explicitly.  Of course, we understand that it 
is non-zero and satisfies the restriction for the reality of the action.

By using the preSUSY invariance, we may derive the corresponding
Ward-Takahashi identities, exactly.  For example, a plaquette-plaquette
correlation is equal to some fermion-fermion correlations.

Obviously there are open questions in our formalism: 1) the relation
between the continuum SUSY and the local fermionic symmetry reported in
this paper; 2) the recovery of the spinor structure for 
the real staggered fermion  $\psi_n$; 3) we also need some explanation of the 
peculiar lattice structure.
Although we have not reached the complete solution to all of these
problems, let us emphasize that it is this lattice structure which
allows us to have the fermionic symmetry.  This is very important.
Because, a naive attempt to construct a model over the whole lattice often
encounters the situation that a fermionic symmetry tends to an $O(a)$
symmetry in a naive continuum limit.  We have avoided this uninteresting
situation owing to the lattice structure chosen in this paper.  The
details on this point will be reported in our forthcoming
paper\cite{sawanaka2}.  

As the reader may have noticed, the interaction in our model goes off
when we remove the fermions.  In the paper\cite{sawanaka2}, we extend
our consideration in the present paper and show that another model may
be constructed naturally which does not have this peculiar nature.

\acknowledgments{This work is supported in part by the Grants-in-Aid for
Scientific Research No. 12640258, 12640259, 13135209, 12640256, 12440060
and 13135205 from the Japan Society for the Promotion of Science.}


\end{document}